# Relationship among Structural, Disordered, Magnetism and Band Topology in MnSb$_2$Te$_4$·(Sb$_2$Te$_3$)$_n$ Family

Ming Xi[1,2,3,#], Yuchong Zhang[4,5,#], Wenju Zhou[6,#], Famin Chen[4,5,#], Donghan Jia[6], Huiyang Gou[6,*], Tian Qian[4,5,*], and Hechang Lei[1,2,*]

[1]School of Physics and Beijing Key Laboratory of Optoelectronic Functional Materials & MicroNano Devices, Renmin University of China, Beijing 100872, China

[2]Key Laboratory of Quantum State Construction and Manipulation (Ministry of Education), Renmin University of China, Beijing 100872, China

[3]School of Physical Science and Technology and ShanghaiTech Laboratory for Topological Physics, ShanghaiTech University, Shanghai 201210, China

[4]Beijing National Laboratory for Condensed Matter Physics and Institute of Physics, Chinese Academy of Sciences, Beijing 100190, China

[5]University of Chinese Academy of Science, Beijing 100049, China

[6]Center for High Pressure Science and Technology Advanced Research, Beijing 100193, China

[#]These authors contributed to this work equally.

[*]Correspondence should be addressed to H. Y. G. (huiyang.gou@hpstar.ac.cn); T. Q. (tqian@iphy.ac.cn); H. C. L. (hlei@ruc.edu.cn).

**Abstract**

Interplay between topology and magnetism induces various exotic quantum phenomena, with magnetic topological insulators (MTIs) serving as a prominent example due to their ability to host the quantum anomalous Hall effect (QAHE). However, the realization of QAHE at higher temperature approaching magnetic-transition-temperature remains a significant challenge, primarily due to the scarcity of suitable material platforms and limited understanding of the intricate relationships between band topology, magnetism, and defects. Here, we report a comprehensive investigation of $MnSb_2Te_4 \cdot (Sb_2Te_3)_n$ ($n$ = 0 - 5) single crystals, including the discovery of novel $MnSb_8Te_{13}$ pure phase. Experimental measurements confirm that $MnSb_8Te_{13}$ exhibits ferromagnetism and features topologically nontrivial electronic structures, characterized by a Dirac point located further from the conduction band and a possible larger bulk gap compared to $MnBi_2Te_4 \cdot (Bi_2Te_3)_n$ ($n$ = 0 - 3). Moreover, we systematically analyze the relationship between structure, magnetism, topology, and disorder within Mn(Sb, Bi)$_2$Te$_4 \cdot$((Sb, Bi)$_2$Te$_3)_n$ family. Present work will shed light on the exploration of potential platforms capable of achieving QAHE near magnetic transition temperature, offering new directions for advancing topological quantum materials.

Since achieving quantum anomalous Hall effect (QAHE) through van-Vleck type magnetic interaction in topological insulators (TIs) is proposed theoretically [1], magnetic-element-doped TIs (Sb, Bi)$_2$(Se, Te)$_3$ have been extensively studied [1-6] and eventually the QAHE is realized at $T \sim 30$ mK in Cr$_{0.15}$(Bi$_{0.1}$Sb$_{0.9}$)$_{1.85}$Te$_3$ [6]. Because the strong magnetic disorder and the presence of localized states introduce additional dissipative transport channels in magnetically doped TIs, however, the temperature of QAHE is typically one to two orders of magnitude ($T \sim$ mK) lower than the magnetic transition temperature $T_C$ [4, 7]. In recent years, a breakthrough has been achieved with the discovery of intrinsic magnetic TI (MTI) MnBi$_2$Te$_4$, enabling higher-temperature QAHE at $T \sim 1$ - $2$ K [8-10]. Moreover, the difference between the temperature realizing QAHE ($T \sim 1.9$ K) and $T_C$ ($\sim 13$ K) has been significantly reduced in MnBi$_{10}$Te$_{16}$ and it also exhibits near-QAHE phenomena up to $T \sim 7$ K [11].

On the other hand, for magnetic-element-doped TIs, many of experimental results demonstrate that at same doping level, the $T_C$ for Sb$_2$Te$_3$ is usually higher than that in Bi$_2$Te$_3$ [7, 12]. It can be understood by the enhanced magnetic interaction in Sb$_2$Te$_3$ when compared to Bi$_2$Te$_3$. In addition, the bulk gap of Sb$_2$Te$_3$ ($\sim 0.26$ eV) [13] is significantly larger than that of Bi$_2$Te$_3$ ($\sim 0.15$ eV) [14]. Thus, it naturally asks whether the MnSb$_2$Te$_4$·(Sb$_2$Te$_3$)$_n$ family has a higher magnetic transition temperature and larger bulk band gap than the MnBi$_2$Te$_4$·(Bi$_2$Te$_3$)$_n$ family, which will help to realize QAHE at higher temperature?

Recently, MnSb$_2$Te$_4$ has attracted intensive attention because it can exhibit the antisite-defect-dependent magnetic and topological properties with much higher $T_C$ than MnBi$_2$Te$_4$ [15-17]. In contrast, the study on MnSb$_2$Te$_4$·(Sb$_2$Te$_3$)$_n$ ($n \geq 1$) is still scarce [18-21]. For the recently reported MnSb$_8$Te$_{13}$, it is uncertain whether synthesized MnSb$_8$Te$_{13}$ is pure phase because of the obviously different XRD pattern from that of isostructural MnBi$_8$Te$_{13}$ and the existence of three transition temperatures in magnetic measurements [22]. In this work, we grew MnSb$_2$Te$_4$·(Sb$_2$Te$_3$)$_n$ ($n = 0$ - $5$) single crystals by using the self-flux method, and carried out a detailed study on these crystals, especially discovered MnSb$_8$Te$_{13}$ pure phase single crystal. It is found that the ferromagnetic MnSb$_8$Te$_{13}$ has a topologically nontrivial electronic structure with Dirac points located further away from the bottom of conduction band (~ 180 meV) and possibly a larger bulk band gap than that in MnBi$_2$Te$_4$·(Bi$_2$Te$_3$)$_n$ ($n$ = 0 - 3) family. These features suggest that MnSb$_2$Te$_4$·(Sb$_2$Te$_3$)$_n$ ($n$ = 1 - 5) is a potential material platform for the realization of QAHE near magnetic transition temperature. Furthermore, we analysis the evolution of physical properties of Mn(Sb, Bi)$_2$Te$_4$·((Sb, Bi)$_2$Te$_3$)$_n$ family with increasing $n$ to make a comprehensive understanding of the relationship between magnetism, topological and disorder in these materials.

Single crystals of MnSb$_2$Te$_4$·(Sb$_2$Te$_3$)$_n$ ($n$ = 0 - 5) were grown by using the self-flux method (see detailed sample growth method in Supplemental Note 1 of Supplemental Materials). As shown in Fig. 1(a), the measured X-ray diffraction (XRD) patterns of MnSb$_2$Te$_4$·(Sb$_2$Te$_3$)$_n$ ($n$ = 0 - 5) exhibit a series of (00$l$) diffraction peaks, nearly identical to those in MnBi$_2$Te$_4$·(Bi$_2$Te$_3$)$_n$ ($n$ = 0 - 5) [23] but different from those of Sb$_2$Te$_3$. With the increase of $n$, the angle difference between two adjacent (00$l$) diffraction peaks ($\Delta\sin\theta$) becomes smaller (Fig. 1(b)). According to the Bragg relation $2d\sin\theta = m\lambda$, where $d$ is the distance between the lattice plane, $\theta$ is the incident angle, $\lambda$ is the X-ray wavelength, and $m$ is a integer, when the $d$ is larger, the corresponding $\theta$ will be smaller and thus the $\Delta\sin\theta$ becomes smaller [24-26]. Moreover, the XRD pattern of a MnSb$_8$Te$_{13}$ single crystal is shown in Fig. S1(a) separately and to make the weak peaks more apparent, we plotted the XRD diffraction pattern with a logarithmic scale for MnSb$_2$Te$_4$·(Sb$_2$Te$_3$)$_n$ ($n$ = 3 – 5) as shown in Fig. S2. This allows us to observe the multiple (00$l$) peaks more clearly. Additionally, the Energy-dispersive X-ray spectroscopy (EDX) spectrum and the scanning electron microscope (SEM) image of MnSb$_2$Te$_4$·(Sb$_2$Te$_3$)$_n$ ($n$ = 0 – 5) as shown in Fig. S3 reveal that all single crystals exhibit typical layered structures with high-quality crystallinity and uniform compositions. As depicted in Fig. 1(c), the $\Delta\sin\theta$ decreases with increasing $n$ gradually. In fact, the value of $d\cdot\Delta\sin\theta$ should be a constant proportional to $\lambda$ (see Fig. S4 in Supplemental Materials). On the other hand, the $d$ is proportional to the interlayer distance between the adjacent Mn-Mn layers $d^{inter}_{Mn-Mn}$, thus the smaller $\Delta\sin\theta$ reflects the increased $d^{inter}_{Mn-Mn}$ with larger $n$ (Fig. 1(c)). Furthermore, even the crystal structures of MnSb$_2$Te$_4$·(Sb$_2$Te$_3$)$_n$ ($n$ = 4, 5) are difficult to resolve due to large $c$-axial values by using single crystal XRD, the $d^{inter}_{Mn-Mn}$ values can still be estimated using the relation $d^{inter}_{Mn-Mn}(n = 4, 5) = <d^{inter}_{Mn-Mn}\cdot\Delta\sin\theta>/\Delta\sin\theta(n = 4, 5)$, where $<d^{inter}_{Mn-Mn}\cdot\Delta\sin\theta>$ represents the average of $d^{inter}_{Mn-Mn}\cdot\Delta\sin\theta$ for $n$ = 0 - 3. It can be seen that the $d^{inter}_{Mn-}$ $_{Mn}$ increases nearly linearly with $n$, confirming the validity of calculated $d^{inter}_{Mn-Mn}$ for $n = 4, 5$. It also indicates that MnSb$_2$Te$_4$·(Sb$_2$Te$_3$)$_n$ ($n = 0$ - 5) exhibits almost the same structural evolution of isostructural MnBi$_2$Te$_4$·(Bi$_2$Te$_3$)$_n$ ($n = 0$ - 5) family. Fig. 1(c) also shows the schematic drawings of MnSb$_2$Te$_4$·(Sb$_2$Te$_3$)$_n$ ($n = 0$ - 5), which are formed by stacking one MnSb$_2$Te$_4$ septuple layers (SL) and $n$ layers of Sb$_2$Te$_3$ quintuple layers (QL) along the $c$-axis alternatively. For the newly discovered MnSb$_8$Te$_{13}$, we solved its crystal structure using the single-crystal XRD (Table S1 - S2 and Fig. S1(b) in Supplemental Materials). It is found that MnSb$_8$Te$_{13}$ is isostructural to MnBi$_8$Te$_{13}$ [25] with the lattice parameters $a = b = 4.2601(6)$ Å and $c = 132.424(19)$ Å with space group $R\bar{3}m$ (No. 166).

MnSb$_8$Te$_{13}$ exhibits a ferromagnetic (FM) transition at $T_C \sim 11.3$ K (Fig. 2(a)), which is determined from the peak of d$\chi$/d$T$ curve (inset of Fig. 2(a)). The bifurcation behavior of zero-field cooling (ZFC) and field-cooling (FC) $\chi(T)$ curves at $\mu_0 H = 0.01$ T should be due to the existence of multi-domain state, which changes into single-domain state at high field. It is noted that MnSb$_8$Te$_{13}$ single crystals show a variation of $T_C$, ranging from 10.5 K - 14 K (Figs. S5 and S6 in Supplemental Materials). Fig. 2(b) shows the field dependence of magnetization $M(\mu_0 H)$ of MnSb$_8$Te$_{13}$ for $H//c$ at various temperatures. When $T < T_C$, MnSb$_8$Te$_{13}$ exhibits an obvious hysteresis loop, confirming the FM ground state, and it is consistent with the $\chi(T)$ results. In contrast, at $T > T_C$, the $M(\mu_0 H)$ curves show a linear dependence on field, a typical paramagnetic behavior. Moreover, for $H//ab$, the hysteresis behavior is absent and the saturation field ($\sim 1$ T) is much larger than that for $H//c$ at 2 K (0.05 T) (Fig. S5 in Supplemental Materials). It suggests that the $c$-axis is the magnetic easy axis for MnSb$_8$Te$_{13}$. Magnetic properties of MnSb$_2$Te$_4$·(Sb$_2$Te$_3$)$_n$ ($n = 0$ - 5) series samples at 2 K are shown in Fig. 2(c) and Figs. S5 - S12 in Supplemental Materials. MnSb$_2$Te$_4$·(Sb$_2$Te$_3$)$_n$ ($n = 0$ - 2) exhibit antiferromagnetic (AFM) ground states while the $n \geq 2$ samples show FM ground states. It is noted that using different synthesis conditions both AFM and FM MnSb$_6$Te$_{10}$ can be obtained. For AFM samples ($n \leq 2$), their initial spin-flop field $\mu_0 H_{sp}$ decrease with increasing $n$ ($\mu_0 H_{sp} \sim 0.45$ T, 0.15 T and 0.016 T for MnSb$_2$Te$_4$, MnSb$_4$Te$_7$ and MnSb$_6$Te$_{10}$, respectively). For FM samples ($n \geq 2$), the coercive field $\mu_0 H_C$ increases gradually with increasing $n$. Fig. 2(d) shows the initial saturation moments $M_{is}$s of MnBi$_2$Te$_4$·(Bi$_2$Te$_3$)$_n$ ($n$ = 0 - 3) and MnSb$_2$Te$_4$·(Sb$_2$Te$_3$)$_n$ ($n$ = 0 - 5) single crystals [24, 27-39]. The $M_i$s of MnBi$_2$Te$_4$·(Bi$_2$Te$_3$)$_n$ ($n$ = 0 - 3) samples are higher than those of MnSb$_2$Te$_4$·(Sb$_2$Te$_3$)$_n$ ($n$ = 0 - 5) samples generally. Furthermore, the $M_i$s of MnSb$_2$Te$_4$·(Sb$_2$Te$_3$)$_n$ ($n$ = 0 - 5) increases with the increase of $n$ in principle. Moreover, the $M_i$s of AFM and FM Mn(Sb/Bi)$_6$Te$_{10}$ samples are different significantly.

MnSb$_8$Te$_{13}$ (sample 1) shows a metallic behavior at $T > 25$ K and the longitudinal resistivity $\rho_{xx}(T)$ increases slightly when the temperature further decreases to ~ 15 K, possibly due to the enhancement of spin scattering caused by spin fluctuation when the temperature approaches $T_C$. (Fig. 3(a) and inset). The $\rho_{xx}(T)$ decreases dramatically at $T_C$ because of the suppression of spin disorder scattering. At lower temperature, the $\rho_{xx}(T)$ shows an upturn behavior possibly due to weak localization effect (WL) [40], electron-electron interaction [41] and/or Kondo effect [42]. Moreover, the peak of $\rho_{xx}(T)$ near $T_C$ is suppressed at field of 5 T (inset of Fig. 3(a)), originating from the decrease of spin-disorder scattering at external field. As shown in Fig. 3(b), the low-field dependence of anomalous Hall resistivity $\rho_{yx}^{AH}(\mu_0H)$ below $T_C$ exhibits similar behavior to the $M(\mu_0H)$ curves (Fig. 2(b)), both of which show obvious hysteresis behaviors. With increasing $T$, the values of $\rho_{yx}^{AH}(\mu_0H = 0.05$ T) decrease gradually and the hysteresis behavior disappear above $T = 12$ K, consistent with the $M(\mu_0H)$ measurements and indicating MnSb$_8$Te$_{13}$ enter pramagnetic state. The high-field $\rho_{yx}^{AH}(\mu_0H)$ curves show positive slopes, indicating the hole-type carriers are dominant (Figs. S13 in Supplemental Materials). The temperature dependence of carrier concentration $n(T)$ and mobility $\mu(T)$ obtained from the linear fits using the high-field $\rho_{yx}^{AH}(\mu_0H)$ curves ($\mu_0H > 3.5$ T) is shown in Fig. 3(c). The $n(T)$ shows a weak temperature dependence, reflecting the dominance of single hole-type carriers. When $T$ is close to $T_C$, the $n(T)$ exhibits a drop, which may be related to the evolution of electronic structure caused by FM transition. On the other hand, the $\mu(T)$ decreases with increasing $T$ in principle due to the increase of various scattering processes such as electron-phonon and/or electron-electron ones.

As shown in Fig. 3(d), the magnetoresistance MR [= $(\rho_{xx}(\mu_0H) - \rho_{xx}(0))/\rho_{xx}(0)$] of MnSb$_8$Te$_{13}$ sample 1 presents a butterfly shape at low-field region (±0.05 T) and 2 K, in agreement with the hysteresis behaviors of $\rho_{yx}^{AH}(\mu_0H)$ and $M(\mu_0H)$ curves. In the field range of ±0.4 T, the MR curve at $T$ = 2 K shows a weak field dependence and some samples can even display a "V" shape (inset of Fig. 3(d)). Moreover, at higher field, there is a negative MR. With the increase of $T$ approaching $T_C$, the negative MR becomes smaller and when $T$ is far above $T_C$, the MR becomes positive (noted as "ordinary positive MR") due to Lorentz force induced by magnetic field. This behavior is distinctively different from the behavior of MR in FM $MnSb_2Te_4$ ($T_C$ = 40 K) (Fig. S12 in Supplemental Materials), in which the negative MR increases when $T$ approaches $T_C$. For the latter, it can be understood by the increase of negative MR (noted as "field-induced negative MR") due to the stronger suppression of spin disorder scattering near $T_C$, as observed in conventional FM materials, and the decrease of ordinary positive MR, which is usually proportional to $(\mu_0H)^2$ [43]. In order to understand such unusual behaviors of MR in $MnSb_8Te_{13}$, we propose a tentative explanation including weak anti-localization (WAL) and weak localization (WL) effects (Figs. 3(e) and 3(f)). The WAL effect can originate from either the strong spin-orbit coupling (SOC) in the bulk materials or spin momentum locking in the topological surface states (SSs) of TIs [44, 45]. It can lead to a quick increased positive MR at low-field region, which saturates at high-field region. On the other hand, the WL effect exists in magnetic-elements-doped TIs commonly, competing with WAL effect in general [46]. It will result in a negative MR which increases with field gradually. The total contributions of WAL and WL effects can lead to the "V" shape MR at low-temperature and low-field region [47]. In addition, if the total MR is positive and increases with temperature quickly, it will cancel out the field-induced negative MR partially. In this case, the total MR will be negative but decreases with temperature gradually. According to the theory proposed by Lu *et al.* [46], the total contributions of WAL and WL effects depend on $\Delta/2E_F$, where $\Delta$ is the energy gap of SS and $E_F$ is the Fermi energy level. In order to realize positive MR, the $\Delta/2E_F$ needs to be significantly less than 0.5 at low temperature. Moreover, the $\Delta$ should decrease with increasing temperature and the $E_F$ should be almost unchanged below $T_C$. Consequently, the decrease of $\Delta/2E_F$ will lead to a larger positive MR at higher temperature [46]. Above analysis suggests that MnSb$_8$Te$_{13}$ should have a topological nontrivial electronic structure with a SS which is gapped by ferromagnetism [48]. Similar MR behaviors have also been observed in MnSb$_6$Te$_{10}$ that is proposed to be a MTI [21]. In contrast, such contributions from WAL and WL effects are absent in topological trivial MnSb$_2$Te$_4$ without SS [17].

The existence of topological nontrivial electronic structure in MnSb$_8$Te$_{13}$ is further confirmed by angle-resolved photoemission spectroscopy (ARPES) measurements, as shown in Fig. 4. The bulk valence band crosses $E_F$ forming a hole-type FS which exhibits a three-fold rotation symmetry (Figs. 4(a) - 4(c)), inconsistent with the six-fold symmetry of the surface Brillouin zone, possibly as a result of the matrix element effect. The bulk conduction band and bulk gap cannot be observed, indicating the hole doping nature of this material. Since the topological property is best demonstrated by the SSs in the bulk gap, we studied the band structure above $E_F$ using time-resolved ARPES, which is based on pump-probe method [49] (Figs. 4(d) - 4(f)). A cut passing through Γ point (Fig. 4(d)) reveals the energy of bulk conduction band and the existence of SS. The bottom of unoccupied bulk conduction band locates at ∼ 530 meV above $E_F$, whereas the unoccupied part of the bulk valence band cannot be observed. More importantly, an SS with Dirac-cone like dispersion can be discerned from the spectrum and the Dirac point locates at ∼ 350 meV above $E_F$. The unoccupied bulk conduction band and the linear dispersion feature of the topological SSs can also be seen in the curvature along EDC and the MDC waterfall plot (Figs. 4(d) and 4(e)). The Dirac velocity of the SS is estimated to be ∼ $4.6 \times 10^5$ m/s, while in Sb$_2$Te$_3$ it is ∼ $3.5 \times 10^5$ m/s [50], and the distance of the Dirac point from the minimum of conduction band ∼ 180 meV is greater than that of Sb$_2$Te$_3$, which is ∼ 120 meV [50-52]. Moreover, no bulk valence band was found below the minimum of conduction band in the range of far beyond ∼ 260 meV (the bulk band gap of Sb$_2$Te$_3$) [13,50-53]. Those pronounced difference and characteristics demonstrates that the SS observed here originates from the topological nature of MnSb$_8$Te$_{13}$, instead of Sb$_2$Te$_3$ multiple layers arising from stacking disorder. It clearly demonstrates the nontrivial topological property of MnSb$_8$Te$_{13}$, which is well consistent with the results of transport measurements. In addition, the results of transport measurement for MnSb$_8$Te$_{13}$ are also significantly different from Sb$_2$Te$_3$. The MnSb$_8$Te$_{13}$ samples are expected to contain four different kind of cleavage terminations with magnetic and nonmagnetic layers. In our experiments—where we performed multiple cleaves and took measurements at different sample regions after each cleavage—only two distinct band-structure variants were observed (see discussion in Supplemental Note 3 of Supplemental Materials). This discrepancy could be attributed to the following three possible effects: large light spot of *tr*-ARPES system ($\geqslant$ 23 μm) [49]; lower energy resolution than conventional ARPES; preferential yield of specific terminations by the cleavage process. It is noted that the energy gap of SS expected to be induced by the long-range *c*-axial magnetic order in MTI [48] and experimentally deduced from transport measurements is not observed below the magnetic transition temperature by ARPES. This could be attributed to four reasons: the magnetic moments at top surface may be still disordered at lowest measuring temperature (8 K), and thus the energy gap of the SS cannot be opened [30]; the electron temperature is higher than the $T_C$ because of the pump pulse; or at the measuring temperature close to $T_C$, the energy gap could be very small, exceeding the limit of energy resolution [49]; the measured terminations have gapless surface states, similar to some QL$_n$ terminations of MnBi$_8$Te$_{13}$.

After providing the experimental evidence of possible nontrivial topological electronic structure of FM MnSb$_8$Te$_{13}$, we try to provide an analysis to establish relationship between structure, antisite disorder, magnetism and band topology in Mn(Sb, Bi)$_2$Te$_4$·((Sb, Bi)$_2$Te$_3$)$_n$ family. The total content of Mn (Mn$_{total}$), the content of Mn at Mn site (Mn$_{Mn}$) and Mn at Sb/Bi site (Mn$_{Sb/Bi}$) (Tabs. S4 in Supplemental Materials) as a function of $T_{C/N}$ of Mn(Sb, Bi)$_2$Te$_4$·((Sb, Bi)$_2$Te$_3$)$_n$ is plotted in Figs. 5(a) - 5(c), respectively [24,27-39]. As shown in Fig. 5(a), first, for the topologically trivial MnSb$_2$Te$_4$, the $T_C$ or AFM Néel temperature $T_N$ exhibits a near-linear dependence on Mn$_{total}$ (blue line) while the $T_{C/N}$ of topologically nontrivial MnSb$_2$Te$_4$ film deviates from this trend obviously (red dot). Second, for topologically nontrivial AFM MnBi$_2$Te$_4$ to MnBi$_4$Te$_7$, the $T_N$ rapidly decreases from ~ 25 K to ~ 12 K (black symbols). In contrast, although the $T_N$ of AFM MnSb$_2$Te$_4$ (~ 19 K) is significantly lower than that of MnBi$_2$Te$_4$, the $T_N$ of MnSb$_4$Te$_7$ (~ 14 K) is higher than that of MnBi$_4$Te$_7$ (green symbols). Third, for Mn(Sb, Bi)$_2$Te$_4$·((Sb, Bi)$_2$Te$_3$)$_n$ (1 ≤ $n$ ≤ 3), the $T_{C/N}$s are insensitive to $n$ (~ 10.5 - 14 K), possibly ascribed to the diminished interlayer coupling (single-layer limit) [54]. Surprisingly, for MnSb$_2$Te$_4$·(Sb$_2$Te$_3$)$_n$ ($n$ = 4, 5), the $T_C$s increase slightly with $n$.

Next, we propose a phenomenological explanation in order to understand above magnetic behaviors. The $T_{C/N}$ is mainly determined by intralayer and interlayer interactions ($J_{//}$ and $J_{\perp}$) with $T_{C/N} = a|J_{//,\text{eff}}|/(b + \log(J_{//,\text{eff}}/J_{\text{eff}}))$, where $a$ and $b$ are constants of the order of 1, $J_{\text{eff}}$ is a combination of $J_{\perp}$ (interlayer magnetic coupling) and $K$ (easy-plane anisotropy coefficient), and $J_{//,\text{eff}} = f(\text{Mn}_{\text{Mn}})J_{//}$ and $f(x)$ is a function of the $x(\text{Mn}_{\text{Mn}})$ (see discussion in Supplemental Note 2 of Supplemental Materials). It is noted that the $J_{//}$ and $J_{\perp}$ depend on the strength of SOC-induced band inversion (van-Vleck-type interaction), the interaction strength between carrier concentration and local magnetic moment (Ruderman-Kittel-Kasuya-Yosida (RKKY)-type interaction), and the distributions of Mn.

First, for topologically trivial MnSb$_2$Te$_4$, previous studies find that the higher the Mn$_{\text{total}}$ will lead to the higher the antisite defects of Mn$_{\text{Sb}}$ [17, 55], and it will result in a higher hole-type carrier concentration because such kind of defects will introduce hole carries ($p$-type doping). As shown in Fig. 5(d), the $T_{C/N}$ of topologically trivial MnSb$_2$Te$_4$ shows a near-linear dependence on carrier concentration (blue line), implying that the dominant magnetic interaction in these materials should be the RKKY-type. It also explain the remarkable near-linear relationship between $T_{C/N}$ and Mn$_{\text{total}}$. In contrast, the insensitivity of $T_N$ of topologically nontrivial MnBi$_2$Te$_4$ to carrier concentration suggests that the van-Vleck-type magnetic interaction should be dominant (see detailed discussion in Supplemental Note 2 of Supplemental Materials). It is noted that topologically nontrivial MnSb$_2$Te$_4$ film has a ~ 100 % Mn$_{\text{Mn}}$ and low Mn$_{\text{Sb}}$ (~ 3.5 %) (Figs. 5(b) and 5(c)) [16, 17]. Correspondingly, its carrier concentration is very low. In contrast, the $T_C$ is much higher than those of topologically trivial MnSb$_2$Te$_4$ crystals with similar Mn$_{\text{total}}$ (Fig. 5(a)), implying that the van-Vleck-type interaction is dominant, similar to MnBi$_2$Te$_4$.

Second, Bi doping leads to the increase of $T_N$ for Mn(Bi$_{1-x}$Sb$_x$)$_2$Te$_4$ accompanying with the decreased carriers concentration and the increased $x(Mn_{Mn})$ (Figs. 5(b) and 5(e)). It indicates that even the RKKY-type interaction becomes weaker due to the decrease of carrier concentration, the enhanced van-Vleck-type interaction because of enhanced SOC by Bi doping and the increase of $J_{//,eff}$ with increasing $x(Mn_{Mn})$ will result in a significant increase of $T_N$ with Bi doping. For the rapid decrease of $T_N$ from MnBi$_2$Te$_4$ to MnBi$_4$Te$_7$, it can not be ascribed solely to the decrease of $J_\perp$ with increasing $n$, and the change of $J_{//,eff}$ which is much larger than $J_\perp$ [56] with $n$ should also play an important role. Because the $J_{//,eff}$ depends on Mn$_{Mn}$ ($f(x)$) and $J_{//}$ simultaneously, even $J_{//}$ is unchanged with $n$, the much lower Mn$_{Mn}$ of MnBi$_4$Te$_7$ than that of MnBi$_2$Te$_4$ (Fig. 5(d)) can still lead to the decease of $T_N$ further. For Mn(Bi$_{1-x}$Sb$_x$)$_2$Te$_4$·((Bi$_{1-x}$Sb$_x$)$_2$Te$_3$)$_n$ ($n$ = 1 - 2), $T_N$ is insensitive to doping (carrier concentration and type) in the full solid solution range ($0 \leq x \leq 1$) (Fig. 5(e)) [57-63]. It implies that there is a band inversion in MnSb$_4$Te$_7$ and MnSb$_6$Te$_{10}$. Furthermore, the $x(Mn_{Mn})$ of MnSb$_2$Te$_4$·(Sb$_2$Te$_3$)$_n$ ($n \geq 1$) is usually lower than those of MnBi$_2$Te$_4$·(Bi$_2$Te$_3$)$_n$ with similar $T_{C/N}$, suggesting that the $J_{//}$ of MnSb$_2$Te$_4$·(Sb$_2$Te$_3$)$_n$ should be larger than those in MnBi$_2$Te$_4$·(Bi$_2$Te$_3$)$_n$. Such larger $J_{//}$ could originate from the larger SOC due to the better alignment of spin and orbit (larger $L \cdot S$) in MnSb$_2$Te$_4$·(Sb$_2$Te$_3$)$_n$ than that in MnBi$_2$Te$_4$·(Bi$_2$Te$_3$)$_n$, even the latter contains heavier Bi element (see discussion in Supplemental Note 2 of Supplemental Materials).

Third, the insensitivity of $T_{C/N}$ for Mn(Sb, Bi)$_2$Te$_4$·((Sb, Bi)$_2$Te$_3$)$_n$ ($1 \leq n \leq 3$) can be explained by the negligible $J_\perp$ and almost unchanged $J_{//,eff}$ with weak $n$ dependence of $x(Mn_{Mn})$ when $n \geq 2$. Moreover, the increase of $T_N$ for MnSb$_2$Te$_4$·(Sb$_2$Te$_3$)$_n$ ($n \geq 4$) could be ascribed to the slight enhancement of $J_{//,eff}$. Because the $x(Mn_{Mn})$ is commonly lower than 1, it implies that the $T_{C/N}$ at single-layer limit (large $n$) could be increased further if the $x(Mn_{Mn})$ can be enhanced.

Finally, we discuss the evolution of band topology in MnSb$_2$Te$_4$·(Sb$_2$Te$_3$)$_n$ ($n$ = 0 - 5) with $n$ and antisite defects (Figs. 5(b) and 5(g)). For both AFM and FM MnSb$_2$Te$_4$, the much higher Mn$_{Sb}$ content (~ 13 %) than that of MnBi$_2$Te$_4$ (~ 3.5 %) could decrease the SOC in Sb$_2$Te$_3$ QL, resulting in a topologically trivial electronic structure. The Mn$_{Sb}$ content decreases rapidly with increasing $n$ in MnSb$_2$Te$_4$·(Sb$_2$Te$_3$)$_n$ and the Mn$_{Sb}$ contents are lower than ~ 3.5 % when $n \geq 2$ (Fig. 5(c)). Thus, it is expected that MnSb$_2$Te$_4$·(Sb$_2$Te$_3$)$_n$ ($n \geq 2$) are topologically nontrivial, confirmed by ARPES measurements for MnSb$_8$Te$_{13}$ (Fig. 4). Meanwhile, the dominant magnetic interaction changes from RKKY-type to van-Vleck-type (Fig. 5(g)). Importantly, the Dirac point of MnSb$_8$Te$_{13}$ is far away below the bottom of conduction band (~ 180 meV) (Fig. 4), and Mn(Sb, Bi)$_2$Te$_4$·((Sb, Bi)$_2$Te$_3$)$_n$ usually have similar or larger bulk band gaps $E_g$ when compared to corresponding (Bi, Sb)$_2$Te$_3$ (for Sb$_2$Te$_3$, $E_g$ ~ 260 meV) [8, 13, 14, 57, 64, 65], thus it is reasonable to speculate that MnSb$_8$Te$_{13}$ may have a larger $E_g$ with Dirac point away from bulk bands when compared to MnBi$_2$Te$_4$·(Bi$_2$Te$_3$)$_n$ because $E_g$(Sb$_2$Te$_3$) > $E_g$(Bi$_2$Te$_3$). As schematically shown in Fig. 5(f), such large band gap with the decreased antisite defects can diminish the localized states and mobility edge in MnSb$_2$Te$_4$·(Sb$_2$Te$_3$)$_n$ ($n \geq 2$) and is favorable for quantum transport phenomena of SSs realized at the temperature close to the $T_{C/N}$. Therefore, MnSb$_2$Te$_4$·(Sb$_2$Te$_3$)$_n$ ($n \geq 2$) may be a more promising material platform than MnBi$_2$Te$_4$·(Bi$_2$Te$_3$)$_n$ to achieve QAHE at higher temperature near $T_C$.

In summary, MnSb$_2$Te$_4$·(Sb$_2$Te$_3$)$_n$ ($n$ = 0 - 5) single crystals are grown by using the self-flux method. Combined transport, magnetization and ARPES measurements confirm that the novel MnSb$_8$Te$_{13}$ has a topologically nontrivial electronic structure and ferromagnetic ground state. Furthermore, we discuss the relation between structure, magnetism, topology and Mn element distribution in Mn(Sb, Bi)$_2$Te$_4$·((Sb, Bi)$_2$Te$_3$)$_n$ in detail and propose that MnSb$_2$Te$_4$·(Sb$_2$Te$_3$)$_n$ family may be a superior platform to study the various quantum phenomena in intrinsic MTIs at elevated temperature.

## Supporting Information

The Methods, magnetic properties, detailed compositions, field dependence of anomalous Hall resistivity and magnetoresistance of MnSb$_2$Te$_4$·(Sb$_2$Te$_3$)$_n$ crystals, the discussion of magnetic interactions in Mn(Sb, Bi)$_2$Te$_4$·((Sb, Bi)$_2$Te$_3$)$_n$ family.

K.; Wang, L. L.; Chen, X.; Xue, Q. K.; Ma, X. C.; and Wang, Y. Y. Band structure engineering in (Bi$_{1-x}$Sb$_x$)$_2$Te$_3$ ternary topological insulators. *Nat. Commun.* **2011,** *2,* 574.

## Acknowledgment

We thank the fruitful discussion with Kai Liu and Ningning Zhao. This work is supported by the National Key R&D Program of China (Grant No. 2022YFA1403800 and 2023YFA1406500), the National Natural Science Foundation of China (12274459, U22A6005), the Chinese Academy of Sciences (XDB33020100), the Synergetic Extreme Condition User Facility (SECUF), H.Y.G. acknowledges the financial support of the National Science Fund for Distinguished Young Scholars (Grant No. T2225027).

## Author contributions

H.C.L. provided strategy and advice for the research; M.X. and H.C.L. performed the crystal growth, magnetization and transport measurements and fundamental data analysis; W.J.Z., D.H.J. and H.Y.G. performed the single-crystal XRD measurements Y.C.Z., F.M.C. and T.Q. performed the ARPES measurements; All authors contributed to the preparation and revisions of the manuscript.

## Author Information

The authors declare no competing financial interests. The data that support the findings of this study are available from the corresponding author H.C.L. (hlei@ruc.edu.cn) upon reasonable request.

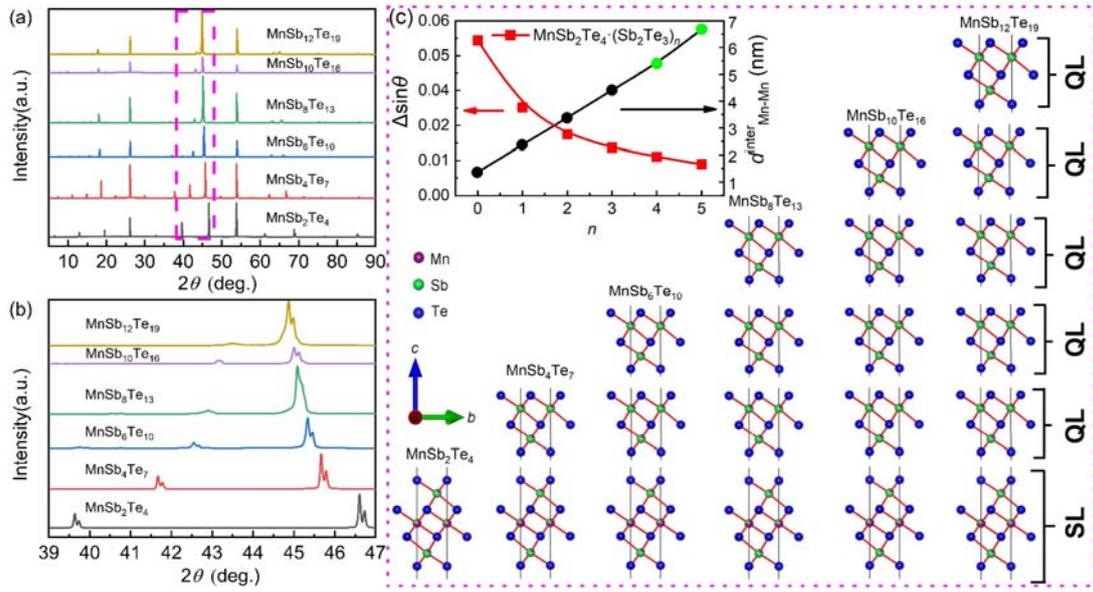

**Figure 1. Crystal structures of MnSb$_2$Te$_4$·(Sb$_2$Te$_3$)$_n$ ($n$ = 0 - 5).** (a) XRD patterns of MnSb$_2$Te$_4$·(Sb$_2$Te$_3$)$_n$ ($n$ = 0 - 5) single crystals. (b) the enlarged view of XRD pattern in the range of 39° - 47° labelled by the magenta rectangular in (a). The asterisk symbols indicate that the $d^{inter}_{Mn-Mn}$ for the compounds with $n$ = 4 and 5 are calculated by using the differences of neighboring diffraction angles of (00$l$) (see main text). (c) Schematic drawings of crystal structures of MnSb$_2$Te$_4$·(Sb$_2$Te$_3$)$_n$ ($n$ = 0 - 5) family with antisite defects between Mn and Sb sites. The purple, green, and blue balls represent Mn, Sb, and Te atoms, respectively. The SL and QL layers represent the MnSb$_2$Te$_4$ septuple layers and Sb$_2$Te$_3$ quintuple layers, respectively. Inset: the angle difference between adjacent peaks in XRD patterns shown in (b) ($\Delta\sin\theta$) and the interlayer distance between the adjacent Mn-Mn layers ($d^{inter}_{Mn-Mn}$) as a function of $n$.

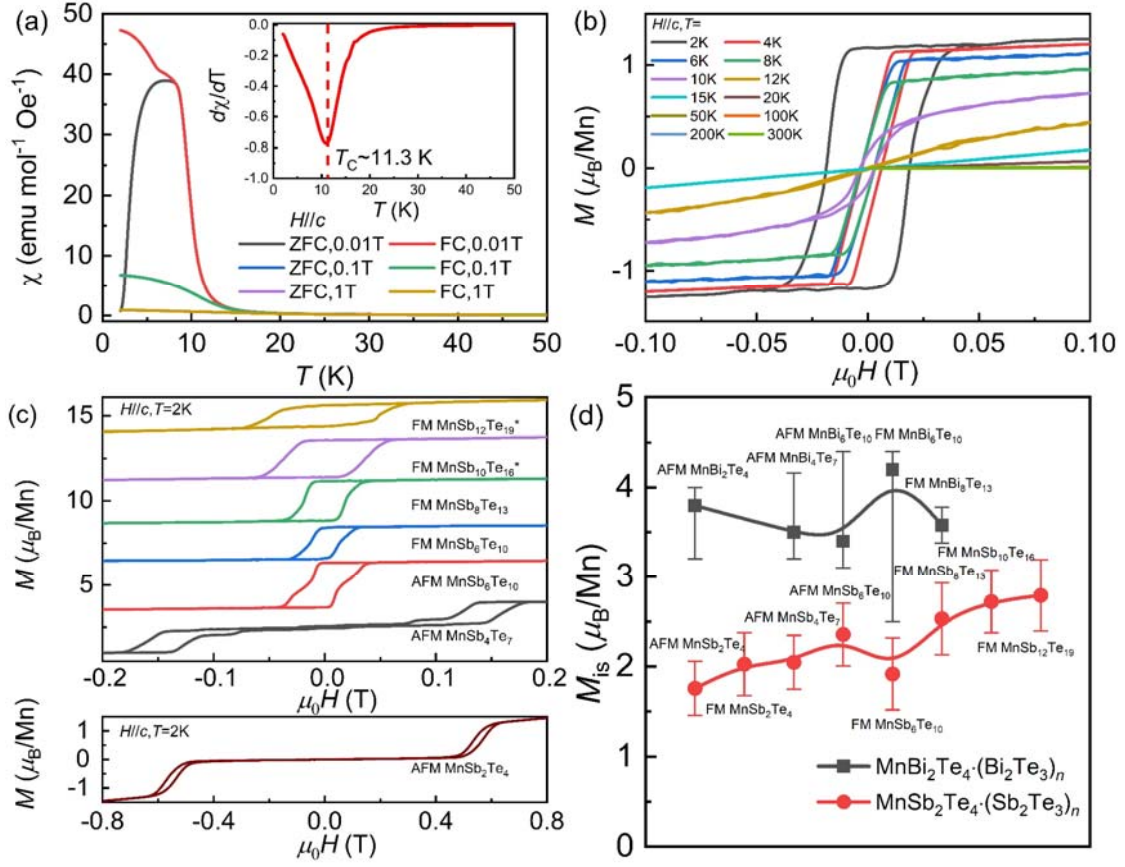

**Figure 2. Magnetic properties of MnSb$_2$Te$_4$·(Sb$_2$Te$_3$)$_n$ ($n$ = 0 - 5) single crystals.** (a) Temperature dependence of magnetic susceptibility $\chi(T)$ of MnSb$_8$Te$_{13}$ measured at $\mu_0 H$ = 0.01 T, 0.1 T and 1 T for $H//c$ with ZFC and FC modes. The inset shows the d$\chi$/d$T$ with $\mu_0 H$ = 0.01 T as a function of temperature at $T \leq 50$ K. (b) Field dependence of magnetization $M(\mu_0 H)$ of MnSb$_8$Te$_{13}$ at different temperatures for $H//c$. (c) Field dependence of $M(\mu_0 H)$ of AFM MnSb$_2$Te$_4$·(Sb$_2$Te$_3$)$_n$ ($n$ = 0 - 2) and FM MnSb$_2$Te$_4$·(Sb$_2$Te$_3$)$_n$ ($n$ = 2 - 5) single crystals at 2 K for $H//c$. (d) The $M_{is}$s of MnBi$_2$Te$_4$·(Bi$_2$Te$_3$)$_n$ ($n$ = 0 - 3) and MnSb$_2$Te$_4$·(Sb$_2$Te$_3$)$_n$ ($n$ = 0 - 5) single crystals. The error bars for MnBi$_2$Te$_4$·(Bi$_2$Te$_3$)$_n$ ($n$ = 0 - 3) represent the maximum and minimum reported values in the literature while the error bars for MnSb$_2$Te$_4$·(Sb$_2$Te$_3$)$_n$ ($n$ = 0 - 5) originate from the uncertainty of samples mass [24, 27-39].

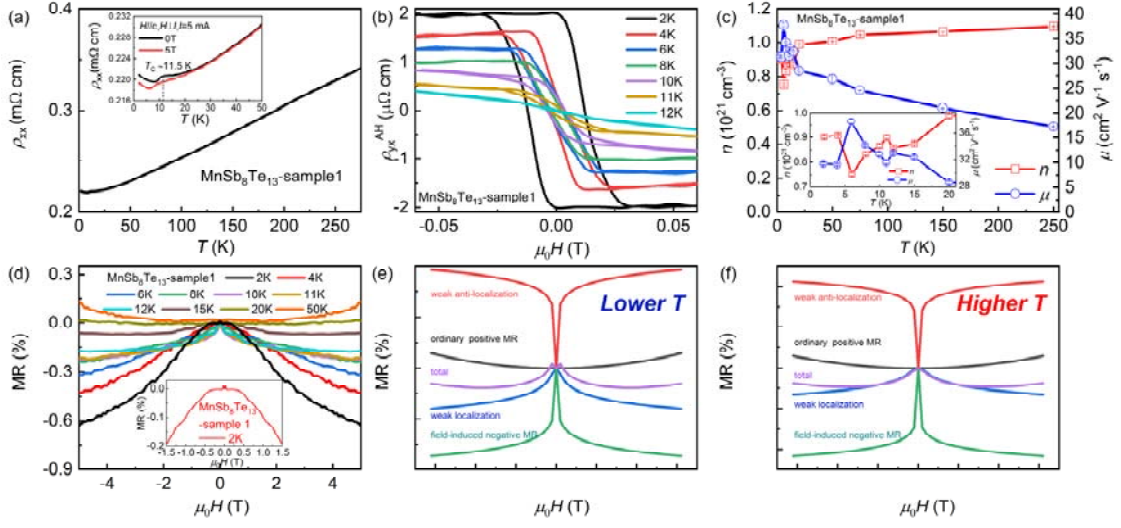

**Figure 3. Electrical properties of MnSb$_8$Te$_{13}$ (sample 1).** (a) Temperature dependence of $\rho_{xx}(T)$ at zero field. The inset shows the $\rho_{xx}(T)$ curves at zero field and 5 T at $T \leq 50$ K. (b) Field dependence of $\rho_{yx}^{AH}(\mu_0 H)$ in the field up to ±0.06 T at various temperatures. (c) Derived $n(T)$ (red squares) and $\mu(T)$ (blue circles) as a function of temperature using the single-band model. The inset shows the enlarged view of $n(T)$ and $\mu(T)$ in the range of 2 K $\leq T \leq$ 20 K. (d) Field dependence of MR in the field up to ±5 T at various temperatures. (e) and (f) Schematic plots of field dependence of MR when considering different contributions at different temperatures below $T_C$.

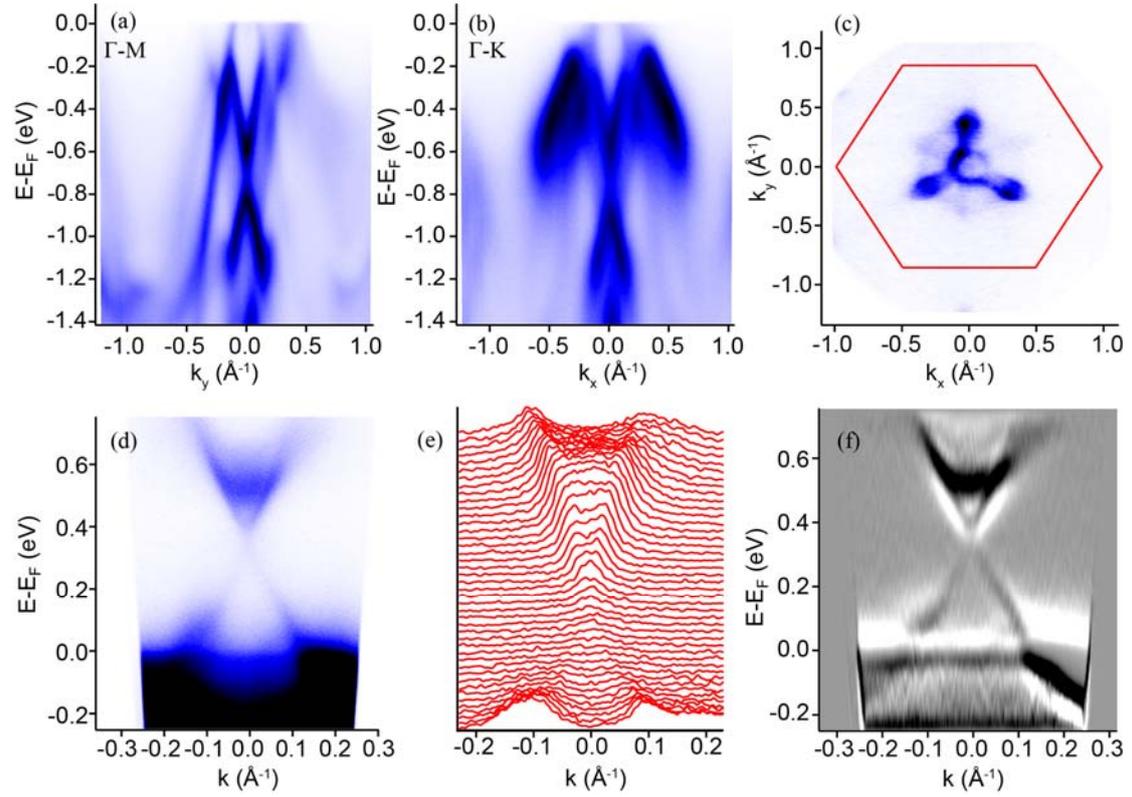

**Figure 4. Electronic structure of MnSb$_8$Te$_{13}$.** Electronic structure along (a) Γ-M (b) Γ-K direction and (c) Fermi surface mapping measured with synchrotron ARPES at $hv$ = 67 eV at 24.7 K. (d) Electronic structure of unoccupied states measured with pump-probe method at $hv$ = 1.2 eV for pump and $hv$ = 7.2 eV for probe at 8 K. (e) Momentum distribution curve (MDC) corresponding to the intensity map in (d), but shown as a log scale plot. (f) Curvature plot along EDC direction of the intensity map in (d).

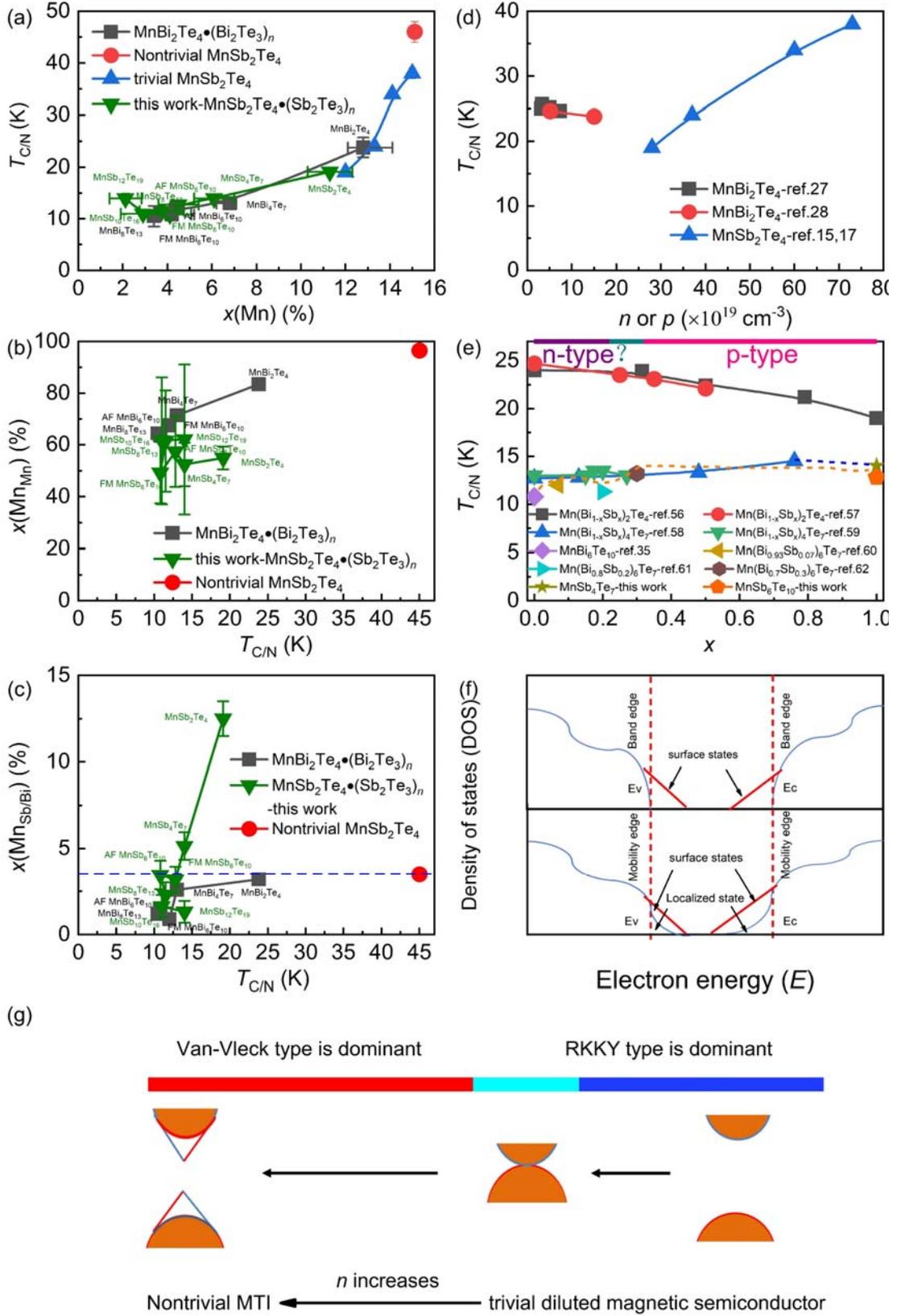

**Figure 5. Evolution of topological and magnetic interaction in Mn(Sb, Bi)$_2$Te$_4$·((Sb, Bi)$_2$Te$_3$)$_n$.** (a) - (c) The Mn$_{total}$, Mn$_{Mn}$ and Mn$_{Sb(Bi)}$ vs. $T_{C/N}$ of Mn(Sb, Bi)$_2$Te$_4$·((Sb, Bi)$_2$Te$_3$)$_n$ [24,27-39]. (d) The carrier concentration ($n$ or $p$) dependence of $T_{C/N}$ for MnBi$_2$Te$_4$ and MnSb$_2$Te$_4$. (e) The dependence of $T_{C/N}$ on $x$ and carrier type of Mn(Bi$_{1-x}$Sb$_x$)$_2$Te$_4$·((Bi$_{1-x}$Sb$_x$)$_2$Te$_3$)$_n$ ($n$ = 0 - 2). (f) Schematic diagram of electron state densities of intrinsic MTI with and without disorder. (g) Diagram of evolution of the electronic band structure of MnSb$_2$Te$_4$·(Sb$_2$Te$_3$)$_n$ ($n$ = 0 - 5) from topologically trivial diluted magnetic semiconductor to nontrivial intrinsic MTI as $n$ increases.